\documentclass[twocolumn,showpacs,preprintnumbers,amsmath,amssymb,nofootinbib,APS]{revtex4}

\usepackage{dcolumn}
\usepackage{bm}

\begin{document}

\preprint{}%gr-qc/xxxxxxx}

\title{ {\bf Post-Newtonian constraints on $f(R)$ cosmologies in Palatini formalism }}
\author{Gonzalo J. Olmo}\email{gonzalo.olmo@uv.es}
\affiliation{ {\footnotesize Departamento de F\'{\i}sica Te\'{o}rica and
    IFIC, Centro Mixto Universidad de
    Valencia-CSIC} \\
    {\footnotesize Facultad de F\'{\i}sica, Universidad de Valencia, Burjassot-46100, Valencia,
    Spain}\\
         {\footnotesize  Physics Department, University of
Wisconsin-Milwaukee,Milwaukee, WI 53201, USA }}

\date{May 26th, 2005}

\begin{abstract}
We compute the complete post-Newtonian limit of the Palatini form
of $f(R)$ gravities using a scalar-tensor representation. By
comparing the predictions of these theories with laboratory and
solar system experiments, we find a set of inequalities that any
lagrangian $f(R)$ must satisfy. The constraints imposed by those
inequalities allow us to find explicit functions that bound from
above and from below the possible nonlinear terms of the
lagrangian. We conclude that the lagrangian $f(R)$ must be almost
linear in $R$ and that nonlinear corrections that grow at low
curvatures are incompatible with observations. This result shows
that modifications of gravity at very low cosmic densities cannot
be responsible for the observed cosmic speed-up.

\end{abstract}


\maketitle

\section{Introduction}

The physical mechanism that drives the recently observed
accelerated expansion rate of the universe \cite{Tonry03,Knop03}
is yet unclear. Among other possibilities, it has been suggested
that it could be due to modifications of the gravitational
interaction at very low cosmic curvatures \cite{CDTT,VOL} (see
also \cite{CAPO}). Since the addition of positive powers of the
scalar curvature to the Hilbert-Einstein lagrangian may lead to
early-time inflation, it is tempting to introduce corrections that
grow at low curvatures to see if the resulting cosmological models
can in some way justify the current observations. The theories
constructed in this way and, with more generality, all theories
where the gravity lagrangian is an arbitrary function of the
scalar curvature, are known as $f(R)$ gravities. They were
originally formulated within the standard metric formalism and
then exported to the Palatini formalism, where metric and
connection are seen as independent fields. In this work we will
study the Palatini formulation of such theories. The metric
formalism will be considered elsewhere \cite{PI}. \\

It has been shown that in the Palatini formalism many different
$f(R)$ lagrangians may lead to late-time cosmic acceleration
\cite{VOL,M-W-Modified,matching-data}. This is true even in cases
in which the function $f(R)$ is completely different from the
Hilbert-Einstein lagrangian, i.e., when $f(R)$ is not just a
perturbation of the linear lagrangian. Though much work has been
carried out recently to understand the cosmological aspects of
these theories, very little is known about their properties in
other regimes. The solar system, where the observational data are
extensive and very precise, represents a more suitable and cleaner
scenario for testing different aspects of the gravitational
interaction than the cosmological regime, where much ``dirty''
astrophysics and uncertain data may be involved. In fact, if in
addition to a modified gravitational dynamics, sources of dark
energy were acting in the cosmic expansion, it would be very
difficult to distinguish one effect from the other. In the solar
system, on the contrary, the matter sources are well known and,
therefore, the gravitational dynamics manifests in its own, with
negligible ``dark'' interferences.\\

This work is thus aimed to analyze in detail the predictions of
the Palatini form of $f(R)$ gravities in the weak-field,
slow-motion or post-Newtonian regime. The Newtonian limit of these
theories has been studied previously in \cite{BAR1}, for the
particular lagrangian $f(R)=R+\omega _0R^2$, and more recently in
\cite{M-W} and \cite{BAR2}, where a general $f(R)$ lagrangian was
considered. Our work will try to go a step further obtaining the
complete post-Newtonian metric and placing explicit constraints on
the form of the lagrangian. In the computation of the
post-Newtonian metric, we will use a scalar-tensor representation
of these theories. Rather than an unnecessary complication or a
mathematical trick, the scalar-tensor form will be very useful in
the interpretation of the equations of motion and the
identification of the matter and the geometrical terms. This
identification is necessary in order to carry out correctly the
post-Newtonian expansion, as we will
point out at due time. \\

The paper is organized as follows. We first derive the equations
of motion and show how to obtain the scalar-tensor representation
out of them. Then we point out some details about the
post-Newtonian expansion, compute the lowest-order corrections of
the metric and discuss the results (the complete post-Newtonian
metric is given in the Appendix). Eventually, we will study the
constraints on the lagrangian $f(R)$ that follow from laboratory
and solar system observational data and will comment on their
consequences in the cosmic regime. We end the paper with a brief
summary and conclusions.

\section{Equations of motion}

The action that defines $f(R)$ gravities in the Palatini formalism
is the following
\begin{equation}\label{eq:def-f(R)}
S[f;g,\Gamma ,\psi_m]=\frac{1}{2\kappa ^2}\int d^4
x\sqrt{-g}f(R)+S_m[g_{\mu \nu},\psi_m]
\end{equation}
where $S_m[g_{\mu \nu },\psi_m]$ represents the matter action,
which depends on the metric $g_{\mu \nu }$ and the matter fields
$\psi_m$. The scalar  $R$ is defined as the contraction $R=g^{\mu
\nu }R_{\mu \nu }(\Gamma )$, where $R_{\mu \nu }(\Gamma )$ has the
form of the Ricci tensor
\begin{equation}\label{eq:def-Ricci}
R_{\mu\nu}(\Gamma )=-\partial_{\mu}
\Gamma^{\lambda}_{\lambda\nu}+\partial_{\lambda}
\Gamma^{\lambda}_{\mu\nu}+\Gamma^{\lambda}_{\mu\nu}\Gamma^{\rho}_{\rho\lambda}-\Gamma^{\lambda}_{\nu\rho}\Gamma^{\rho}_{\mu\lambda}
\end{equation}
and $\Gamma^\alpha_{\beta \gamma }$ is an affine connection
independent of $g_{\mu \nu }$. Varying eq.(\ref{eq:def-f(R)}) with
respect to the metric we obtain
\begin{equation}\label{eq:met-var}
f'(R)R_{\mu\nu}(\Gamma )-\frac{1}{2}f(R)g_{\mu\nu}=\kappa ^2T_{\mu
\nu }
\end{equation}
where $f'(R)\equiv df/dR$. Note that the trace of this equation
\begin{equation}\label{eq:trace}
f'(R)R-2f(R)=\kappa ^2T
\end{equation}
implies an algebraic relation between $R$ and $T$ more involved
than that existing in General Relativity (GR), say, $R=\kappa
^2T$. We will denote by $R(T)$ the algebraic solution to
eq.(\ref{eq:trace}).\\

The variation of the action with respect to the connection gives
\begin{equation}\label{eq:con-var}
\nabla_\lambda  \left[\sqrt{-g}\left(\delta ^\lambda _\alpha
f'g^{\beta \gamma }-\frac{1}{2}\delta ^\beta _\alpha f'g^{\lambda
\gamma }-\frac{1}{2}\delta ^\gamma  _\alpha f'g^{\beta
\lambda}\right)\right]=0
\end{equation}
where $f'=f'(R(T))$ is also an algebraic function of the matter
terms. Using an auxiliary tensor $t_{\mu \nu }=f'g_{\mu \nu }$,
eq.(\ref{eq:con-var}) can be readily solved \cite{MTW}. The
solution states the compatibility between the connection
$\Gamma^\alpha_{\beta \gamma }$ and the {\it metric} $t_{\mu
\nu}$. In other words, $\Gamma^\alpha_{\beta \gamma }$ can be
written as the Levi-Civita connection of $t_{\mu \nu}$
\begin{equation}\label{eq:Gamma-1}
\Gamma^\alpha_{\beta \gamma }=\frac{t^{\alpha \lambda
}}{2}\left(\partial_\beta t_{\lambda \gamma }+\partial_\gamma
t_{\lambda \beta }-\partial_\lambda t_{\beta \gamma }\right)
\end{equation}
Inserting this solution for $\Gamma^\alpha_{\beta \gamma }$,
written in terms of $g_{\mu \nu }$ and $f'$, in
eq.(\ref{eq:met-var}) we obtain
\begin{eqnarray}
R_{\mu \nu }(g)-\frac{1}{2}g_{\mu \nu }R(g)&=&\frac{\kappa
^2}{f'}T_{\mu \nu }-\frac{R(T)f'-f}{2f'}g_{\mu \nu
}-\nonumber\\&-&\frac{3}{2(f')^2}\left[\partial_\mu f'\partial_\nu
f'-\frac{1}{2}g_{\mu \nu }(\partial f')^2\right]+\nonumber \\
&+& \frac{1}{f'}\left[\nabla_\mu \nabla_\nu f'-g_{\mu \nu }\Box
f'\right] \label{eq:Gmn}
\end{eqnarray}
where $R_{\mu \nu }(g)$ and $R(g)$ are computed in terms of the
Levi-Civita connection of the metric $g_{\mu \nu }$, i.e., they
represent the usual Ricci tensor and scalar curvature. To make our
notation clearer, since $t_{\mu \nu }$ and $g_{\mu \nu }$ are
conformally related, it follows that $R(T)=g^{\mu \nu }R_{\mu \nu
}(\Gamma )$ and $R(g)=g^{\mu \nu }R_{\mu \nu }(g)$ are related by
\begin{equation}
R(T)=R(g)+\frac{3}{2f'}\partial_\lambda f'\partial^\lambda
f'-\frac{3}{f'}\Box f'
\end{equation}
where, recall, $f'=f'(R(T))$ is a function of $T$.\\

Introducing the following definitions
\begin{eqnarray}\label{eq:phi=f'}
\phi&\equiv& f'\\
V(\phi)&\equiv& R(\phi)f'-f(\phi) \label{eq:V=rf'-f}
\end{eqnarray}
where we have algebraically inverted $f'(R)$ to obtain $R=R(f')$,
we can write eq.(\ref{eq:Gmn}) in the scalar-tensor form
\begin{eqnarray}\label{eq:Gab}
G_{\mu \nu}({g})&=&\frac{\kappa ^2}{\phi}{T}_{\mu \nu}-
\frac{V}{2\phi}{g}_{\mu \nu } -\nonumber\\&-&
\frac{3}{2\phi^2}\left(\partial_\mu \phi\partial_\nu  \phi -
\frac{1}{2}{g}_{\mu \nu }({\nabla}
\phi)^2\right)+\\
&+&\frac{1}{\phi}\left({{\nabla}}_\mu  {{\nabla}}_\nu
\phi-{g}_{\mu \nu }{\Box}\phi\right) \nonumber
\end{eqnarray}
 The scalar field $\phi $ satisfies
eq.(\ref{eq:trace}), which in the new notation turns into
\begin{equation}\label{eq:field}
2V-\phi V'=\kappa ^2T
\end{equation}
where $V'\equiv dV/d\phi $. This representation of the theory
turns out to be equivalent to an $\omega =-3/2$ Brans-Dicke-like
scalar-tensor theory. For more details on this scalar-tensor
representation and a different derivation of this
result see \cite{P0}.\\

According to eq.(\ref{eq:field}), the scalar field can be
algebraically solved as $\phi =\phi (T)$, i.e., it is a
non-dynamical field. Thus, the effect of the nonlinear lagrangian
in the Palatini formalism is rather different from its effect in
the metric formalism \cite{PI}. In the metric case, the
compatibility between metric and connection gives rise to
additional degrees of freedom in the theory, which can be encoded
in a dynamical scalar field. In the Palatini formalism, however,
the independent connection keeps the order of the equations of
motion and modifies the way matter generates space-time curvature.
In other words, the right hand side of eq.(\ref{eq:Gab})
represents a generalized energy-momentum tensor of matter in which
the trace $T$ plays an enhanced role by means of the terms $\phi
=\phi (T)$ and its derivatives.

\section{Post-Newtonian metric}
The complete post-Newtonian limit needs the different components
of the metric  evaluated to the following orders $g_{00}\sim
O(2)+O(4)$, $g_{0j}\sim O(3)$, $g_{ij}\sim O(2)$ (see
\cite{WILL}). For convenience, we will discuss here only the
lowest-order corrections, $g_{00}\sim O(2)$, $g_{ij}\sim O(2)$,
which will be enough to place important constraints on the gravity
lagrangian. The details of the calculations can be found in the
Appendix. In our calculations, we will use coordinates in which
the outer regions of the local system are in free fall with
respect to the surrounding cosmological model. This guarantees
that the local metric can be made Minkowskian at a given distance
far from the system and fixes all the boundary conditions
necessary for this problem. Thus, we will compute the metric as a
perturbation around the Minkowski background, i.e., $g_{\mu \nu
}\approx \eta_{\mu \nu }+h_{\mu \nu }$. Since the scalar field is
non-dynamical, i.e., it is determined by the local matter
distribution, no boundary conditions need to be set for it.  On
the other hand, once the solution $\phi =\phi (T)$ is obtained, it
could be expanded to different orders of approximation in the
post-Newtonian expansion using the fact that for a perfect fluid
$T=-\rho (1+\Pi -3P/\rho )\approx-\rho +\rho O(v^2/c^2)$, where
$\rho $ is the rest-mass density, $\Pi $ is the specific energy
density (ratio of energy density to rest-mass energy), and $P$ is
the pressure (see chapter 4 of \cite{WILL}). In this way one would
obtain an expansion of the form $\phi (T)\approx \phi (-\rho
)+\partial_T \phi|_{-\rho} O(v^2/c^2)+\ldots $ However, this is an
unnecessary complication of our notation and, therefore, we will
keep $\phi(T) $ exact in our calculations. Note that this
expansion in post-Newtonian orders is different from an expansion
around the vacuum $\phi(T) \approx \phi (0)+\partial_T \phi(0)
T+\ldots$ such as the one apparently considered in \cite{M-W} and
\cite{BAR2} using the original $f(R)$ representation. In their
calculations they expanded the function $f(R)$ around a de Sitter
background characterized by a constant curvature $R_0$. The fact
that in the Palatini approach $R=R(T)$ (do not confuse $R(T)$ with
$R(g)$), implies that $R_0=R(T=0)$. Thus, an expansion of $f(R)$
around $R_0$ actually represents an expansion around $T\approx 0$,
which is an expansion around the vacuum, not an expansion in
post-Newtonian orders. Since the functional dependence of $f(R) $
with $T$ is a priori unknown, there is no guarantee that such
expansion around the vacuum can be valid in the range from $T=0$
 up to the typical densities inside planets, stars or laboratory-size
bodies. In fact, the weak-field slow-motion limit does not require
low densities but not too high matter concentrations and low
matter velocities, $v^2/c^2\ll 1$. Thus, the conclusions regarding
the Newtonian limit obtained in \cite{M-W} and \cite{BAR2} could
not be valid. This point will be clarified below in detail. In our
description in terms of a scalar field, $\phi (T)\equiv f'[R(T)]$,
the role of the $\phi (T)$ terms is clear from the very beginning:
they represent new contributions of the matter sources to the
equations of motion. In consequence, they must be treated as
matter terms and expanded in
post-Newtonian orders, not around the vacuum. \\

\subsection{Second-order corrections}\label{sec:A}
For convenience,  we introduce a dimensionless quantity
$\tilde{\phi }=\phi/\phi_0$, where $\phi _0\equiv \phi (0)$ is the
vacuum reference value, and define $\Omega(T)
\equiv\log(\tilde{\phi })$. In order to get $h_{ij}$ diagonal and
to respect the perturbative description, we find that $\Omega(T) $
must be seen, at least, of order $O(v^2)$. We will indicate with a
superindex the order of approximation of each quantity when
necessary. To second order, the metric satisfies the equations
\begin{eqnarray}
-\frac{1}{2}\nabla^2\left[h_{00}^{(2)}-\Omega ^{(2)}\right]&=&
\frac{\kappa ^2\rho-V(\phi )}{2\phi}\\
-\frac{1}{2}\nabla^2\left[h_{ij}^{(2)}+\delta _{ij}\Omega
^{(2)}\right]&=& \left[\frac{\kappa ^2\rho+V(\phi
)}{2\phi}\right]\delta _{ij}
\end{eqnarray}
where we have used the gauge condition\footnote{ This condition
was already used in \cite{BAR1}, and also in \cite{BAR2}. In their
notation, $-\partial_k\Omega=b_k$.} $h^\mu _{k,\mu
}-\frac{1}{2}h^\mu _{\mu ,k}=\partial_k\Omega$. These equations
admit the following solutions
\begin{eqnarray}\label{eq:h00}
h_{00}^{(2)}(t,x)&=& \frac{\kappa ^2}{4\pi \phi _0}\int d^3x'
\frac{\left[\rho(t,x') -V(\phi )/\kappa ^2\right]}{\tilde{\phi
}|x-x'|}+\Omega ^{(2)}\\
h_{ij}^{(2)}(t,x)&=& \left[\frac{\kappa ^2}{4\pi \phi _0}\int
d^3x' \frac{\left[\rho(t,x') +V(\phi )/\kappa
^2\right]}{\tilde{\phi }|x-x'|}-\Omega ^{(2)}\right]\delta
_{ij}\label{eq:hij}
\end{eqnarray}
In these equations, the local term $\Omega ^{(2)}=\log[\tilde{\phi
}(-\rho )]$ represents a new effect that is not present in the
general Brans-Dicke-like case $\omega \neq -3/2$ (see \cite{PI}).
The contribution due to $\Omega $ is identically zero only if
$f(R)$ is linear (GR and GR plus cosmological constant) and,
therefore, its presence would imply the nonlinearity of the
gravity lagrangian. It is worth noting that rather than an
integrated quantity (cumulative effect), it is directly related to
the local matter density. In consequence, an isolated body will
contribute to the exterior space-time metric
 by means of the integral terms of eqs.(\ref{eq:h00}) and
(\ref{eq:hij}) only. If we now put an object in orbit around the
first one, the metric at the position of this new body will be
modified by the local term $\Omega $ and by the self-gravity of
the body. For the moment, we will concentrate on the integral
terms of eqs.(\ref{eq:h00}) and (\ref{eq:hij}) (isolated body).\\

Assuming that the main contribution to the metric in the solar
system is due to the  sun, we can express eqs.(\ref{eq:h00}) and
(\ref{eq:hij}) outside the sun as follows
\begin{eqnarray}\label{eq:sun}
h_{00}^{(2)}(t,x)&=&2G\frac{M_\odot}{r}-\frac{V_0}{\phi _0}\frac{r^2}{6}\\
h_{ij}^{(2)}(t,x)&=&\left[2\gamma
G\frac{M_\odot}{r}+\frac{V_0}{\phi _0}\frac{r^2}{6}\right]\delta
_{ij}
\end{eqnarray}
In these expressions,  $G$ and $\gamma $ are defined as
\begin{eqnarray}\label{eq:G-Pal}
G&=&\frac{\kappa ^2}{8\pi \phi_0}\left(1+\frac{M_V}{M_\odot}\right)\\
\gamma &=&\frac{M_\odot-M_V}{M_\odot+M_V}\label{eq:g-Pal}
\end{eqnarray}
where $M_\odot\equiv \int d^3x' \rho (t,x')/\tilde{\phi }$,
$M_V\equiv \kappa ^{-2}\int d^3x' [V_0-V(\phi)/\tilde{\phi}]$ and
$V_0=V(\phi _0)$. Since the cosmological constant term $V_0/\phi
_0$ must be negligible in solar system scales in order not to
affect the local dynamics, we find a constraint on the function
$f(R)$. We need to note that the value $\phi _0$ is solution of
eq.(\ref{eq:field}) with $T=0$. Using that equation and the
definition of $V(\phi )$ in terms of $f(R)$, it follows that
$V_0=f(R_0)$, where $R_0$ is solution of eq.(\ref{eq:trace})
outside the sun, i.e., $R_0=R(T=0)$. From these considerations it
follows that
\begin{equation}\label{eq:V0}
\left|\frac{f(R_0)}{f'(R_0)}\right|L_L^2\ll 1
\end{equation}
where $L_L$ represents a (Large) length scale the same order or
greater than the solar system and $R_0$ presumably is of order the
cosmological constant $\Lambda  \sim 10^{-53}$m$^{-2}$. \\

Let us consider now the observational constraints on $G$ and
$\gamma $. It is well known that in dynamical scalar-tensor
theories the effective $G$ and $\gamma $ depend on two cosmic
parameters, say, the state of the field, $\phi _0$, and the range
$m_\varphi^{-1} $ of its interaction, which are the same for all
bodies \cite{PI,WILL,WAG}. In the non-dynamical situation
discussed here, $G$ and $\gamma $ are not universal quantities,
i.e., they are not the same for all bodies. According to the
definitions given above after eq.(\ref{eq:g-Pal}), two bodies with
the same $M_\odot$ do not necessarily have the same value $M_V$
and, therefore, may lead to different values of $G$ and $\gamma $.
This is due to the fact that $M_\odot$ and $M_V$ are defined as
integrals over quantities related to $\phi =\phi (T)$, whose
values depend on the structure and composition of the body.
Obviously, the experimental evidence supporting the universality
of $G$ and the measurements of $\gamma \approx 1$ \cite{WIL-liv}
indicate that $|M_V/M_\odot|\ll 1$. The only cases in which
$M_V=0$ correspond to GR and GR plus cosmological constant, i.e.,
those cases in which the lagrangian $f(R)$ is linear, or
$V=V_0=$constant. All nonlinear lagrangians predict a non trivial
potential $V(\phi )$ and, therefore, a non vanishing $M_V$, which
may give rise to the effects discussed above. Unfortunately, the
fact that $M_V$ is given as an integrated quantity does not allow
us to place any explicit constraint on the form of the function
$f(R)$. On the other hand, it is quite disturbing the fact that a
body with Newtonian mass $M_N\equiv \int d^3x' \rho (t,x')$ may
yield different values of $M_\odot$, $G$ and $\gamma $ depending
on its internal properties. Stated another way, a given amount of
Newtonian mass may lead to gravitational fields of different
strengths and dynamical properties. Since, as far as we know,
effects of this type have not been observed in laboratory, we
expect a very weak dependence of $M_\odot$ on $\phi$. This is
equivalent to saying that $\phi$ cannot change too much with the density.\\

It is worth noting that with the definitions given above for $G$
and $\gamma $ and neglecting the cosmological constant term, we
can write for an isolated body $h_{00}^{(2)}\equiv 2U$, where $U$
represents the Newtonian potential. It is thus easy to see that
the term $(h_{00}^{(2)})^2/2$ of the complete post-Newtonian limit
(see the Appendix) leads to the PPN parameter $\beta=1 $, like in
GR. The remaining higher-order terms of the metric are all
affected by $\phi $.

\subsection{ $\Omega ^{(2)}$ contribution}

We will now analyze the effect of the term $\Omega ^{(2)}$ that
 we omitted above in the case of an isolated massive body. As we
pointed out, this term must be taken into account when a test body
is placed within the gravitational field of another body. Thus, it
must be present in any physical situation. Neglecting the
cosmological constant contribution for simplicity, we can write
the metric as follows
\begin{eqnarray}
h_{00}^{(2)}(t,x)&=& 2U(r)+\Omega ^{(2)}(T)\\
h_{ij}^{(2)}(t,x)&=& [2\gamma U(r)-\Omega ^{(2)}(T )]\delta
_{ij}\label{eq:hij-omega}
\end{eqnarray}
where $U(r)\equiv GM_\odot/r$ is the Newtonian potential generated
by the massive body and $\Omega ^{(2)}(T )=\log[\phi(T)/\phi _0]$
is a local term that depends on the matter density $T=-\rho $ at
the point $(t,x)$, where our test body is located. First thing
 we need to note is that $\Omega ^{(2)}(T)$ is a perturbative
quantity of order $O(v^2/c^2)\ll 1$. The only manner to respect
the perturbative approach is accepting that $\phi (T)$ depends
very weakly on $\rho $, i.e., that $\phi (T)$ must be almost
constant over a wide range of densities and can be well
approximated by $\phi (T)=\phi _0+(\partial \phi /\partial
T)|_{T=0} T+\ldots$, with $\phi _0^{-1}(\partial \phi /\partial
T)|_{T=0}T\ll 1$ from $T=0$ up to nuclear densities ($T=10^{14}$
g/cm$^3$) at least. The need for this expansion about $\phi_0$
indicates that the lagrangian must be almost linear in $R$ (recall
that $\phi \equiv df/dR$). Furthermore, if $\phi (T)$ had a
stronger dependence on $T$, individual atoms could experience
strong accelerations due to sudden changes in $\Omega ^{(2)}$ when
going from outside atoms to inside atoms. Those individual
microscopic gravitational effects would manifest in the
macroscopic, averaged, description of matter. Since such effects
have not been observed, they must be very small, if they actually
exist. Thus, the weak dependence of $\phi $ on $T$ within this
wide density interval confirms that the contribution of the
nonlinear terms to the lagrangian $f(R)$ must be very small. This
conclusion agrees with our previous claims regarding
the weak dependence of $M_\odot$, $G$ and $\gamma $ on $\phi (T)$.\\

Let us analyze in detail the dependence of $\phi $ on $T$. We will
consider, as an illustration, the Newtonian limit of the
conservation equations $\nabla_\mu T^{\mu \nu }=0$ of a perfect
fluid. These equations lead to
\begin{eqnarray}
\frac{\partial \rho }{\partial t}&+&\nabla(\rho \vec{v}) = 0\\
\rho \frac{d\vec{v}}{dt}&=& \rho
\nabla\left(\frac{h_{00}^{(2)}}{2}\right)-\nabla P
\end{eqnarray}
where a modification with respect to the classical Euler equations
is introduced by the term $\Omega$ contained in $h_{00}^{(2)}$.
This modification is given by
\begin{equation}
\frac{\rho }{2}\nabla \Omega ^{(2)}=-\frac{\rho
}{2}\frac{(\partial \phi /\partial T)}{\phi }\nabla\rho
\end{equation}
and requires that the condition
\begin{equation}\label{eq:ratio}
\left| \frac{\rho (\partial \phi /\partial T)}{\phi }\right|\ll 1
\end{equation}
be satisfied over the wide range of densities mentioned above in
order to guarantee the validity of the macroscopic classical Euler
equations. Note that eq.(\ref{eq:ratio}) must be true in general,
since the contribution of $\Omega ^{(2)}$ to the acceleration of a
body is given in terms of $\nabla \Omega ^{(2)}$. This constraint
can be rewritten using eq.(\ref{eq:field}) to evaluate $\partial
\phi /\partial T$ as follows
\begin{equation}\label{eq:ratio1}
\left|\frac{(\kappa ^2\rho/\phi ) }{(\phi V''-V') }\right|\ll 1
\end{equation}
It is remarkable the fact that the denominator $[\phi V''-V']$ in
eq.(\ref{eq:ratio1}) is the counterpart of the effective square
mass $m_\varphi ^2\equiv [\phi _0V_0''-V_0']/[3+2\omega ]$
associated to dynamical Brans-Dicke-like fields with $\omega
\neq-3/2$ \cite{PI}. For our discussion it will be more convenient
to see this effective mass as an inverse length defining the
interaction range of the scalar field. We can thus interpret
eq.(\ref{eq:ratio1}) as the quotient of two lengthscales, one
associated to the scalar field over another related to the mass
density, $L^{-2}(\rho )\equiv (\kappa ^2\rho c/\phi_0 )$.
Eq.(\ref{eq:ratio1}) can then be seen as the counterpart of the
condition $m_\varphi ^2 L^2\gg 1$ that corresponds to $\omega \neq
-3/2$ theories. Written in terms of the lagrangian $f(R)$,
eq.(\ref{eq:ratio1}) turns into
\begin{equation}\label{eq:f-const}
R\tilde{f}'(R)\left|\frac{\tilde{f}'(R)}{R\tilde{f}''(R)}-1\right|
L^2(\rho )\gg 1
\end{equation}
where $\tilde{f}'\equiv f'/f'_0=\phi /\phi _0$. According to our
interpretation of the denominator of eq.(\ref{eq:ratio1}), we must
demand that the multiplicative factor in front of $L^2(\rho )$ in
eq.(\ref{eq:f-const}) satisfies
\begin{equation}\label{eq:f-const1}
\left|\frac{\tilde{f}'(R)}{R\tilde{f}''(R)}-1\right|\ge
\frac{1}{l^2 R\tilde{f}'}
\end{equation}
where $l^2$ represents a lengthscale much smaller than $L^2(\rho
)$ at nuclear densities, which is the shortest $L^2(\rho )$ that
we can associate to ordinary matter. This inequality will allow us
to find out which lagrangians satisfy the condition given in
eq.(\ref{eq:f-const}) at densities such that $L^2(\rho )\gg l^2$.
Notice that the modulus of eq.(\ref{eq:f-const1}) may lead to more
than one solution. Let us consider first the case
\begin{equation}\label{eq:f-const+}
\frac{\tilde{f}'(R)}{R\tilde{f}''(R)}-1\ge \frac{1}{l^2
R\tilde{f}'}
\end{equation}
Defining $A\equiv 1/(R\tilde{f}')>0$, eq.(\ref{eq:f-const+}) turns
into
\begin{equation}\label{eq:A+}
-\left[\frac{2+\frac{R}{A}\frac{dA}{dR}}{1+\frac{R}{A}\frac{dA}{dR}}\right]\ge
\frac{A}{l^2}
\end{equation}
Since the right hand side of this equation is positive, the left
hand side must also be positive. This can only happen if the
denominator is negative and the numerator is positive. In fact, if
we denote $-\varepsilon \equiv 1+\frac{R}{A}\frac{dA}{dR}$,
eq.(\ref{eq:A+}) can be written as $(1-\varepsilon )/\varepsilon
\ge A/l^2$. Thus, only if $\varepsilon <1$ the  condition of the
sign can be satisfied. In addition, for a highly linear lagrangian
we expect $A/l^2\gg 1$. This leads to $\varepsilon \ll 1$, which
is compatible with $\tilde{f}'\approx 1$. The sign and magnitude
of the denominator indicate that $\tilde{f}''$ must be very small
and positive. A little algebra is enough to show that
\begin{equation}\label{eq:A+1}
d\log[A(A+2l^2)]\ge d\log[\frac{1}{R^2}]
\end{equation}
Once integrated, the new inequality can be written as
\begin{equation}\label{eq:A+2}
(A-A_+)(A-A_-)\ge 0
\end{equation}
where an integration constant, $c_0^2$, appears in
\begin{equation}\label{eq:A+-}
A_{\pm}=\frac{-l^2R \pm \sqrt{c_0^2+(l^2R)^2}}{R}
\end{equation}
Since $A$ must be positive ($\phi \equiv f'>0$ to have a
well-posed theory), the only valid solution to eq.(\ref{eq:A+2})
is $A\ge A_+$, which implies
\begin{equation}\label{eq:A+3}
0<\frac{df}{dR}\le \sqrt{(f_0')^2+(l^2R)^2}+l^2R
\end{equation}
where we have fixed $c_0=f'_0$ to eliminate the tilde from
$\tilde{f}'$. We can finally integrate this last inequality to
obtain
\begin{equation}\label{eq:A+4}
{f}\le \alpha
+\frac{l^2{R}^2}{2}+\frac{{R}}{2}\sqrt{(f_0')^2+(l^2{R})^2}+
\frac{f_0'}{2l^2}\log[l^2{R}+\sqrt{1+(l^2{R})^2}]
\end{equation}
Before commenting this result, let us consider the second
inequality that follows from eq.(\ref{eq:f-const1})
\begin{equation}\label{eq:f-const-}
\frac{\tilde{f}'(R)}{R\tilde{f}''(R)}-1\le -\frac{1}{l^2
R\tilde{f}'}
\end{equation}
Using again the function $A\equiv 1/(R\tilde{f}')$, we obtain
\begin{equation}\label{eq:A-}
\left[\frac{2+\frac{R}{A}\frac{dA}{dR}}{1+\frac{R}{A}\frac{dA}{dR}}\right]\ge
\frac{A}{l^2}
\end{equation}
This inequality can only be satisfied if
$\left(1+\frac{R}{A}\frac{dA}{dR}\right)>0$, which corresponds to
$\tilde{f}''<0$. Simple algebraic manipulations lead to
\begin{equation}\label{eq:A-1}
\frac{df}{dR}\ge \sqrt{(f_0')^2+(l^2R)^2}-l^2R
\end{equation}
which integrated gives
\begin{equation}\label{eq:A-2}
{f}\ge \alpha
-\frac{l^2{R}^2}{2}+\frac{{R}}{2}\sqrt{(f_0')^2+(l^2{R})^2}+
\frac{f_0'}{2l^2}\log[l^2{R}+\sqrt{1+(l^2{R})^2}]
\end{equation}

Let us discuss now the significance of eqs.(\ref{eq:A+4}) and
(\ref{eq:A-2}). The constant $l^2$ was introduced in
eq.(\ref{eq:f-const1}) to represent the length scale over which
the nonlinear contributions of the gravity lagrangian were
relevant. For a given $l^2$, the nonlinear effects will begin to
be important about a certain high density scale at which $L^2(\rho
)/l^2\sim 1$. If, for instance, we take $l^2=0$, the nonlinear
effects would be completely suppressed, since then
eq.(\ref{eq:f-const})  would be satisfied at all densities. The
choice $l^2=0$ forces the lagrangian to be linear, which can be
seen from eqs.(\ref{eq:A+4}) and (\ref{eq:A-2}) in the limit
$l^2\to 0$. This limit also indicates that the constant $f_0'$ can
be naturally set to unity. On the other hand, if the nonlinear
terms were relevant at low cosmic curvatures, the lengthscale $l$
would be of order the radius of the universe and, therefore, the
nonlinear effects would dominate the gravitational dynamics at all
scales. This fact is obviously in contradiction with our experience,
as we have discussed in detail throughout this work.\\

Though the inequalities derived above are only strictly valid in
the limit of relatively low curvatures, $l^2R\ll 1$ (far from the
early-time inflationary period), eqs.(\ref{eq:A+4}) and
(\ref{eq:A-2}) not only estimate the leading order of the possible
nonlinear corrections, but give precise functions that bound the
nonlinearities of the gravity lagrangian in this limit. Expanding
around $l^2R\ll 1$ we find
\begin{equation}\label{eq:bound}
\alpha+R-\frac{l^2R^2}{2}\le f(R) \le \alpha  +R+\frac{l^2R^2}{2}
\end{equation}
which confirms that the lagrangian is almost linear in $R$ and
that the leading order corrections can grow, as much, quadratically in $R$.\\

\subsection{ On the Newtonian limit }

Before concluding, we will briefly discuss the Newtonian limit
obtained in \cite{M-W} and \cite{BAR2}. As we mentioned above, the
expansion around the vacuum carried out in those papers is not
valid a priori. We have shown, however, that a viable theory must
admit such expansion because of the experimental evidence
supporting the weak dependence of $\phi $ on $T$. This conclusion,
in our case, came out after analyzing  the predictions of the
theory. In \cite{M-W} and \cite{BAR2}, the expansion was due to an
apparent failure to correctly identify the matter terms and the
geometrical terms. In any case, expanding $\Omega ^{(2)}(T)$
around $T=0$ we reproduce the term $A\rho(x) $ obtained in
\cite{BAR2}. This term, however, is not present in \cite{M-W}. On
the other hand, a Yukawa-type exponential correction in the
Newtonian potential was found in \cite{M-W} and \cite{BAR2}, which
is not present in our calculations. In the case of a dynamical
field satisfying a second-order differential equation, a term of
this type is expected to be related to the interaction range of
the field. In the Palatini case, the field is non-dynamical and,
therefore, there is no reason for such a term. Moreover, assuming
spherical symmetry, the Palatini equations admit exact
Schwarzschild-de Sitter solutions
$ds^2=-A(r)dt^2+dr^2/A(r)+r^2d\Omega ^2$ with $A(R)=1-\alpha
/r+\Lambda r^2/3$. The effect of the asymptotic background
curvature is given by the $\Lambda r^2/3$ term, which has the same
form as the cosmological constant term $(V_0/\phi _0)r^2$ that
appears in our eqs.(\ref{eq:h00}) and (\ref{eq:hij}). Thus, there
is no reason to expect an exponential correction related to the
background curvature. The error seems to be due to a failure in
the identification of the leading-order contribution of the term
$(Rf'-f)g_{\mu \nu }/f'$. In our case, this term is represented by
$(V/\phi )g_{\mu \nu }$, and its leading order is $(V/\phi
)\eta_{\mu \nu }$ not $(V/\phi )h_{\mu \nu }$, which could justify
the Yukawa-type correction for $V/\phi =$constant.

\section{Summary and conclusions}

In this work we have computed the complete post-Newtonian metric
of the Palatini form of $f(R)$ gravities using a scalar-tensor
representation of those theories, which corresponds to the case
$\omega =-3/2$ of Brans-Dicke-like scalar-tensor theories. We have
also discussed the experimental constraints on the lowest-order
corrections of the resulting metric. We have found that the
presence of nonlinear terms in the gravity lagrangian makes the
post-Newtonian metric strongly dependent on the local properties
of the gravitating system. In particular, the effective
gravitational mass $M_\odot$, Newton's constant $G$ and the PPN
parameter $\gamma $ of an isolated body may depend on the internal
structure and composition of the body. This follows from the fact
that those quantities are given in terms of integrals that depend
on $V(\phi )$ and $\phi $, which are functions of the trace $T$.
Since those predictions are not compatible with observations, we
have concluded that $\phi (T)$ should depend very weakly on $T$.
This point has been confirmed by the requirement that the
contribution $\Omega^{(2)}(T)$ should be much smaller than unity
in order to respect the perturbative approach. Analyzing the
effect of the $\Omega^{(2)}(T)$ term in the Newtonian limit, we
have found that the nonlinear corrections at relatively low
curvatures ($l^2R\ll 1$) are bounded by eqs.(\ref{eq:A+4}) and
(\ref{eq:A-2}), which are compatible with an almost linear
lagrangian. These results show that $f(R)$ gravities in the
Palatini formalism with nonlinear terms that grow at low
curvatures, such as the Carroll et al. model \cite{CDTT} suggested
in \cite{VOL}, cannot represent valid mechanisms to justify the
cosmic acceleration.  Nonetheless, Palatini models in which the
nonlinear terms are suppressed below the limits imposed by
eqs.(\ref{eq:A+4}) and (\ref{eq:A-2}), may still be compatible
with observations, though they lead to a theory virtually
undistinguishable from General Relativity plus a cosmological
constant.
\section*{Acknowledgements}

I am grateful to Prof. Leonard Parker and Prof. Jos\'{e} Navarro-Salas
for their always bright comments and insights. I also thank O.
Varela, M. Nebot and J.A. Sans for useful discussions. This work
has been supported by a fellowship from the Regional Government of
Valencia (Spain) and the research grant BFM2002-04031-C02-01 from
the Ministerio de Educaci\'{o}n y Ciencia (Spain), and could not have
been carried out without the comprehension of Sonia G.B.

\appendix
\section{Detailed calculations}

The equations of motion for the metric in Brans-Dicke-like
theories are given by
\begin{eqnarray}\label{eq:Rab}
R_{\mu \nu}&=& \frac{\kappa ^2}{\phi }\left[T_{\mu \nu
}-\frac{1}{2}g_{\mu \nu }T\right]+\frac{\omega }{\phi^2}
\partial_\mu \phi
\partial_\nu \phi +\frac{1}{\phi }\nabla_\mu \nabla_\nu \phi
+\nonumber\\&+&\frac{1}{2\phi }g_{\mu \nu }\left[\Box \phi +V(\phi
)\right]
\end{eqnarray}
The expansion of the different components of $R_{\mu \nu }$ can be
found in the Appendix of \cite{PI}. We will just remark here that
in the case $\omega =-3/2$ the scalar field cannot be expanded in
the same manner as in the general case $\omega \neq -3/2$, since
now the field is non-dynamical and is completely determined by the
matter distribution of the local system, $\phi =\phi (T)$. In the
general case, however, the field is a dynamical entity whose state
is determined by the Universe as a whole. The post-Newtonian
system only contributes with local fluctuations from the
background asymptotic state. In the $\omega =-3/2$ case, due to
the fact that $T=-\rho (1+\Pi -3P/\rho )\approx -\rho +\rho
O(v^2)$, we could expand $\phi(T)\approx \phi (-\rho
)+\partial_T\phi \rho O(v^2)$, though this seems an unnecessary
complication. We will keep all $\phi $ terms exact in our
calculations and will expand them at the end up to the necessary
order. According to this, $\tau _{\mu \nu }\equiv \frac{\kappa
^2}{\phi }\left[T_{\mu \nu }-\frac{1}{2}g_{\mu \nu }T\right]$ is
given by
\begin{eqnarray}
\tau _{ij}&=& \frac{\kappa^2\rho }{2\phi}\delta _{ij}+\rho
O(v^2)\\
\tau _{0j}&=&-\frac{\kappa ^2}{\phi}\rho v_j+\rho O(v^3)\\
\tau_{00}&=&\frac{\kappa ^2\rho }{2\phi}\left[1+\Pi
+2v^2-h_{00}^{(2)}+\frac{3P}{\rho }\right]+\rho O(v^4)
\end{eqnarray}
The contribution coming from the scalar field terms can be written
as
\begin{equation}
\tau ^\phi _{\mu \nu }=(\omega +1)\partial_\mu \Omega\partial_\nu
\Omega +\nabla_\mu \nabla_\nu \Omega +\frac{1}{2}g_{\mu \nu
}\left[\frac{V}{\phi }+\Box \Omega +\partial_\lambda
\Omega\partial^\lambda \Omega\right]
\end{equation}
where $\Omega \equiv \log (\phi/\phi _0) $ and $\phi_0$ is an
arbitrary constant that may be fixed as $\phi _0=\phi (T=0)$. The
components of $\tau ^\phi _{\mu \nu }$ are
\begin{eqnarray}
\tau ^\phi _{ij}&=& \partial_i\partial_j\Omega +\frac{\delta _{ij}}{2}\left[\frac{V}{\phi }+\nabla^2\Omega \right] \\
\tau ^\phi _{0j}&=& \partial_0\partial_j\Omega \\
\tau ^\phi _{00}&=&
\frac{3}{2}\ddot{\Omega}+\frac{1}{2}\partial_k\Omega \partial^k
h_{00}^{(2)}-\frac{1}{2}\nabla^2\Omega
+\nonumber\\&+&\frac{1}{2}\left[h^\mu _{k,\mu }-\frac{1}{2}h^\mu
_{\mu ,k}-\partial_k\Omega
\right]\partial^k\Omega -\nonumber\\
&-&\frac{V}{2\phi
}(1-h_{00}^{(2)})+\frac{h^{ij}}{2}\partial_i\partial_j\Omega
+\frac{1}{2}h_{00}^{(2)}\nabla^2\Omega
\end{eqnarray}
where we assume $\Omega \sim O(v^2)$ at least to guarantee
$h_{ij}$ diagonal and a consistent post-Newtonian expansion.\\

Equating the left hand side of eq.(\ref{eq:Rab}) to its right hand
side, a little algebra leads to
\begin{eqnarray}
-\frac{1}{2}\nabla^2[h^{(2)}_{ij}+\delta _{ij}\Omega
]&+&\frac{1}{2}\partial_i\left[h^\mu _{j,\mu }-\frac{1}{2}h^\mu
_{\mu ,j}-\partial_j\Omega \right]+\nonumber\\
&+&\frac{1}{2}\partial_j\left[h^\mu _{i,\mu }-\frac{1}{2}h^\mu
_{\mu ,i}-\partial_i\Omega \right]=\nonumber \\ &=& \frac{\delta
_{ij}}{2\phi}\left[\kappa ^2\rho
+V(\phi )\right]\\
-\frac{1}{2}\nabla^2h^{(3)}_{0j}&+&\frac{1}{2}\partial_j\left[h^\mu
_{0,\mu }-\frac{1}{2}h^\mu _{\mu ,0}-\partial_0\Omega
\right]+\nonumber\\&+& \frac{1}{2}\partial_0\left[h^\mu _{j,\mu
}-\frac{1}{2}h^\mu _{\mu ,j}-\partial_j\Omega \right] =\nonumber \\
&=& -\frac{\kappa ^2 }{\phi }\rho v_j
\end{eqnarray}
\begin{eqnarray}
-\frac{1}{2}\nabla^2\left[h^{(4)}_{00}+\frac{(h^{(2)}_{00})^2}{2}\right]+\frac{1}{2}\nabla^2\Omega
&+&\nonumber \\ +\frac{1}{2}\partial_0\left[h^\mu _{0,\mu
}-\frac{1}{2}h^\mu _{\mu
,0}-\partial_0\Omega+\frac{1}{2}h^{(2)}_{00,0} \right]&+&\nonumber\\
+\frac{1}{2}\left[h^\mu _{k,\mu }-\frac{1}{2}h^\mu _{\mu
,k}\right]\partial^k h^{(2)}_{00}+\frac{1}{2}h^{(2)}_{00}\nabla^2\left[h^{(2)}_{00}-\Omega \right]&=& \nonumber\\
=\frac{\kappa ^2\rho }{2\phi }\left[1+\Pi
+2v^2-h^{(2)}_{00}+\frac{3P}{\rho
}\right]&+&\nonumber\\+\frac{1}{2}\left[h^\mu _{k,\mu
}-\frac{1}{2}h^\mu _{\mu
,k}-\partial_k\Omega\right]\partial^k\Omega&+&\nonumber\\+
\frac{\ddot{\Omega}}{2}+\frac{1}{2}\partial_k\Omega \partial^k
h^{(2)}_{00} -\frac{V}{2\phi}(1-h^{(2)}_{00})&+&\nonumber
\\+\frac{h^{ij}}{2}\partial_i\partial_j \left[h^{(2)}_{00}-\Omega
\right]
\end{eqnarray}
Using the gauge conditions
\begin{eqnarray}
h^\mu _{k,\mu }-\frac{1}{2}h^\mu _{\mu ,k}&=&\partial_k\Omega\\
h^\mu _{0,\mu }-\frac{1}{2}h^\mu _{\mu
,0}&=&\partial_0\Omega-\frac{1}{2}h^{(2)}_{00,0}
\end{eqnarray}
the equations of above become
\begin{eqnarray}
-\frac{1}{2}\nabla^2[h^{(2)}_{ij}+\delta _{ij}\Omega
]&=&\frac{\delta _{ij}}{2\phi}\left[\kappa ^2\rho
+V(\phi )\right]\\
-\frac{1}{2}\nabla^2h^{(3)}_{0j}-\frac{1}{4}h^{(2)}_{00,0j}&=&-\frac{\kappa
^2
}{\phi }\rho v_j\\
-\frac{1}{2}\nabla^2\left[h^{(4)}_{00}-\Omega
+\frac{(h^{(2)}_{00})^2}{2}\right]&=& \frac{\kappa ^2\rho }{2\phi
}\left[1+\Pi +2v^2+\right.\nonumber
\\ &+& \left.h^{(2)}_{[ij]}+\frac{3P}{\rho }\right]+\frac{\ddot{\Omega }}{2} -\nonumber
\\ &-&\frac{V}{2\phi}(1+h^{(2)}_{[ij]})
\end{eqnarray}
where $h_{[ij]}$ denotes the $ij$-component of $h^{(2)}_{ij}$ and
we have used that, to second order, $h^{(2)}_{00}$ satisfies
\begin{equation}
-\frac{1}{2}\nabla^2\left[h^{(2)}_{00}-\Omega
\right]=\frac{1}{2\phi }[\kappa ^2\rho -V(\phi )]
\end{equation}
post-Newtonian corrections to the metric are thus given by (we
denote $\tilde{\phi }\equiv \phi /\phi _0$)
\begin{eqnarray}
h^{(2)}_{ij}(t,x)&=& \left[\frac{\kappa ^2}{4\pi \phi _0}\int
d^3x'\frac{\left[\rho (t,x')+V(\phi )/\kappa
^2\right]}{\tilde{\phi
}|x-x'|}-\right.\nonumber \\ &-& \left. \Omega ^{(2)}\right]\delta _{ij}\\
h^{(2)}_{00}(t,x)&=& \frac{\kappa ^2}{4\pi \phi _0}\int
d^3x'\frac{\left[\rho (t,x')-V(\phi )/\kappa
^2\right]}{\tilde{\phi
}|x-x'|}+\Omega ^{(2)}\\
h^{(3)}_{0j}(t,x)&=&-\frac{\kappa ^2}{2\pi \phi _0}\int d^3x'
\frac{\rho (t,x')v'_j}{\tilde{\phi }|x-x'|}+\nonumber
\\ &+& \frac{1}{8\pi }\int
d^3x' \frac{h^{(2)}_{00,0j}(t,x')}{|x-x'|}\\
h^{(4)}_{00}(t,x)&=& \frac{\kappa ^2}{4\pi \phi _0}\int d^3x'
\frac{\rho (t,x')}{\tilde{\phi }|x-x'|}\left[1+\Pi +2v^2+\right.
\nonumber \\ &+& \left. h^{(2)}_{[ij]}+3P/\rho
\right]-\nonumber\\&-& \frac{\kappa ^2}{4\pi \phi _0}\int d^3x'
\frac{V(\phi
)\left[1+h^{(2)}_{[ij]}\right]+\ddot{\Omega}^{(2)}/2}{\tilde{\phi
}|x-x'|}-\nonumber\\&-& \frac{(h^{(2)}_{00})^2}{2}+\Omega ^{(4)}
\end{eqnarray}

\end{document}